\documentclass[11pt]{article}
\usepackage{graphicx}
\usepackage[margin=1.25in]{geometry}
\usepackage[usenames,dvipsnames]{color}
\usepackage{url}
\usepackage[colorlinks = true,
            linkcolor = blue,
            urlcolor  = blue,
            citecolor = blue,
            anchorcolor = blue]{hyperref}

\newcommand\pubdate{\today}

\def\Title#1{\begin{center} {\LARGE #1 } \end{center}}
\def\Author#1{\begin{center}{ \sc #1} \end{center}}
\def\Address#1{\begin{center}{ \it #1} \end{center}}

\newcommand\pubblock{\rightline{\begin{tabular}{l} 
         \pubdate \end{tabular}}}
\newenvironment{Abstract}{\begin{quotation} \begin{center}
                       ABSTRACT
     \end{center}\bigskip  }{\end{quotation}}

\def\beq{\begin{equation}}
\def\eeq#1{\label{#1}\end{equation}}
\def\eeqn{\end{equation}}

\newenvironment{Eqnarray}%
   {\arraycolsep 0.14em\begin{eqnarray}}{\end{eqnarray}}
\def\beqa{\begin{Eqnarray}}
\def\eeqa#1{\label{#1}\end{Eqnarray}}
\def\eeqan{\end{Eqnarray}}

\let\bar=\overbar

\def\lsim{\mathrel{\raise.3ex\hbox{$<$\kern-.75em\lower1ex\hbox{$\sim$}}}}
\def\gsim{\mathrel{\raise.3ex\hbox{$>$\kern-.75em\lower1ex\hbox{$\sim$}}}}

\def\del{\partial}
\def\Dslash{\not{\hbox{\kern-4pt $D$}}}
\def\dslash{\not{\hbox{\kern-2pt $\del$}}}
\def\pslash{\not{\hbox{\kern-2pt $p$}}}
\def\ETmiss{\not{\hbox{\kern-4pt $E$}}_T}

\def\Dlr{\mathrel{\raise1.5ex\hbox{$\leftrightarrow$\kern-1em\lower1.5ex\hbox{$D$}}}}

\def\MSB{{\bar{M \kern -2pt S}}}
\def\msb{{\bar{\scriptsize M \kern -1pt S}}}

\def\drb{{\bar{\scriptsize D \kern -1pt R}}}

\newcommand\snowmass{\begin{center}\rule[-0.2in]{\hsize}{0.01in}\\\rule{\hsize}{0.01in}\\
\vskip 0.1in Submitted to the  Proceedings of the US Community Study\\ 
on the Future of Particle Physics (Snowmass 2021)\\ 
\rule{\hsize}{0.01in}\\\rule[+0.2in]{\hsize}{0.01in} \end{center}}

\newcommand{\bnl}[1]{#1$^{1}$}
\newcommand{\lbl}[1]{#1$^{2}$}
\newcommand{\fnal}[1]{#1$^{3}$}
\newcommand{\anl}[1]{#1$^{4}$}

\begin{document}

\pubblock

\Title{Portability: A Necessary Approach\break for Future Scientific Software}

\Author{\fnal{Meghna Bhattacharya},
        \lbl{Paolo Calafiura},
        \anl{Taylor Childers},
        \anl{Mark Dewing},
        \bnl{Zhihua Dong},
        \fnal{Oliver Gutsche},
        \anl{Salman Habib},
        \lbl{Xiangyang Ju},
        \fnal{Michael Kirby}, 
        \fnal{Kyle Knoepfel}, 
        \fnal{Matti Kortelainen}, 
        \fnal{Martin Kwok}, 
        \lbl{Charles Leggett}, 
        \bnl{Meifeng Lin}, 
        \bnl{Vincent R. Pascuzzi},
        \fnal{Alexei Strelchenko},
        \bnl{Brett Viren},
        \lbl{Beomki Yeo},
        \bnl{Haiwang Yu}}
\Address{$^{1}$Brookhaven National Laboratory, Upton, NY 11973, USA}
\Address{$^{2}$Lawrence Berkeley National Laboratory, Berkeley, CA 94720, USA}
\Address{$^{3}$Fermi National Accelerator Laboratory, Batavia, IL 60510, USA}
\Address{$^{4}$Argonne National Laboratory, Lemont, IL 60439, USA}

\begin{Abstract}
\noindent Today’s world of scientific software for High Energy Physics (HEP) is powered by x86 code, while the future will be much more reliant on accelerators like GPUs and FPGAs. The portable parallelization strategies (PPS) project of the High Energy Physics Center for Computational Excellence (HEP/CCE) is investigating solutions for portability techniques that will allow the coding of an algorithm once, and the ability to execute it on a variety of hardware products from many vendors, especially including accelerators. We think without these solutions, the scientific success of our experiments and endeavors is in danger, as software development could be expert driven and costly to be able to run on available hardware infrastructure. We think the best solution for the community would be an extension to the C++ standard with a very low entry bar for users, supporting all hardware forms and vendors. We are very far from that ideal though. We argue that in the future, as a community, we need to request and work on portability solutions and strive to reach this ideal.
\end{Abstract}

\snowmass

\def\thefootnote{\fnsymbol{footnote}}
\setcounter{footnote}{0}

\newpage

Today’s world of scientific software for High Energy Physics (HEP) is powered by x86 code. In today’s research environment, code that is written for the x86 platform is pretty much guaranteed to run everywhere in the world, from computing centers using batch systems to our own laptops. Through the proliferation of x86 hardware the challenge to write scientific code is reduced to how well and efficiently the code can be engineered for the platform. 

But we already see signs that the world is changing. On the High Performance Computing level for sure, where large installations of hardware using GPUs and other accelerators provide more processing power for the same energy consumption as with x86-based supercomputers. This is continued even to the hardware in our computers and laptops, that are starting to move to System-on-a-chip (SoC) architectures as lately Apple with the M1 processor. And the increase in heterogeneity does not stop there, multiple different GPU vendors and CPU vendors are available to optimize the hardware to our research problems. 

This makes the challenge of writing efficient scientific code and getting the science out a lot more difficult. It can lead to designing algorithms and implementing them for specific hardware platforms and combinations, and making it very difficult to use not only one but several of these platforms. This adds a new constraint to getting the science out. It is not anymore enough to have access to sufficient computational power, it also needs to be the correct architecture(s) for which the code was developed.

We argue that it would be easier if researchers could develop scientific software and then could execute it on many different hardware combinations without having to rewrite the code over and over again. This white paper is arguing for solutions that enable the writing of portable code by describing one of the current projects to investigate such technologies.

\section{The Portable Parallelization Strategies project}
\begin{figure}
\begin{center}
\includegraphics[width=0.90\hsize]{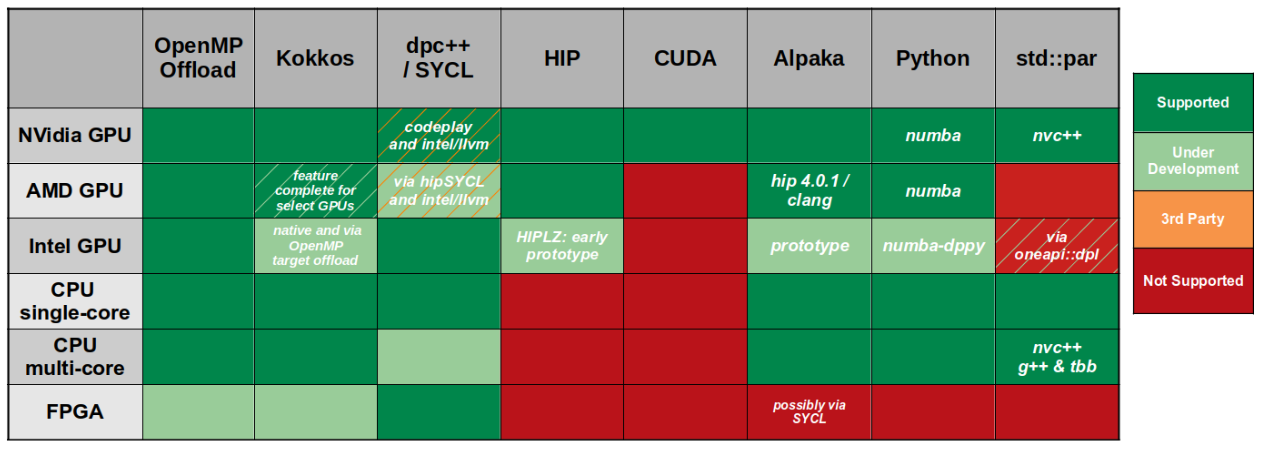}
\end{center}
\caption{Matrix of portability technologies supporting a variety of hardware architectures.}
\label{fig:matrix}
\end{figure}

The High Energy Physics Center for Computational Excellence (HEP/CCE) is a pilot project whose mandate is to provide strategies for HEP experiments to adapt to using increasingly heterogeneous High Performance Computers. It is split into 4 parts, targeting portable parallelization strategies (PPS), fine-grained I/O and related storage issues (IOS), event generators (EG), and complex workflows (CW)~\cite{cce_webpage}. The PPS group is investigating various portability solutions that will permit single source code to be compiled for and executed on multiple different heterogeneous architectures. This is becoming an essential requirement, as each of these architectures use different languages and APIs, such as CUDA~\cite{cuda} for NVIDIA GPUs, SYCL~\cite{sycl} for Intel GPUs, HIP~\cite{hip} for AMD GPUs, and HLS~\cite{Duarte_2018} for FPGAs (see Fig. \ref{fig:matrix}). HEP experiments, which now have code bases in the million lines of source code, do not have the person power to port their CPU code to each back end. Furthermore, validating and maintaining multiple versions of algorithms written in different languages would be exceedingly onerous.

The HEP/CCE-PPS group is currently evaluating Kokkos~\cite{kokkos,CarterEdwards20143202}, SYCL~\cite{sycl}, Alpaka~\cite{MathesP3MA2017,ZenkerAsHES2016,worpitz_2015_49768}, OpenMP/OpenACC~\cite{dagum1998openmp} and {\tt std::execution::parallel}~\cite{std_par} by porting a small number of HEP test beds taken from several different experiments to each portability layer (see Sec. \ref{sec:usecases}). These are
\begin{itemize}
    \item Patatrack and P2R from CMS~\cite{:2008zzk} which perform pixel detector pattern recognition and tracking
    \item The WireCell toolkit from DUNE~\cite{DUNE:2016hlj} which performs space point formation in the liquid argon time projection chamber (TPC)
    \item FastCaloSim from ATLAS~\cite{Aad:2008zzm} which does a fast parameterized simulation of the liquid argon calorimeter
    \item A pixel detector tracking workflow from the “A Common Tracking Software” project (ACTS)~\cite{andreas_salzburger_2022_6220148}
\end{itemize}

The HEP/CCE-PPS group is tightly integrated with a number of HEP experiments, with core developers from each experiment being represented in the group. Each port will be evaluated according to a set of metrics (see Sec.~\ref{sec:metrics}), and at the end of the process, the group will make recommendations back to the experiments and the HEP community in general as to the suitability of each technology that was investigated. It should be noted that no one best solution is likely to exist, as the needs and characteristics of each experiment are different.
\section{Metrics}
\label{sec:metrics}
With the goal to assist the process to make recommendations to HEP experiments and the HEP community, we designed a set of metrics that are of interest to the HEP communities and how scientific software is being developed, and evaluate all portability technologies in question using this set of properties. The metrics set aims at evaluating the whole programming experience for the developer/user using the portability solution, not just the specification or the capability of the solution. Hence we collected HEP use-case programs (see Sec.~\ref{sec:usecases}) for portability solutions from different sub-fields and implement them in different portability technologies. We get hands-on experience for applying a certain technology to HEP software problems by implementing our use cases in the portability technologies under consideration. This includes building, debugging, and adapting existing code to a given technology, which will be reflected in the evaluation of the metrics. 

The metrics set will serve as a point-of-reference for the information about these portability solutions, which more often still lack the needed level of documentation, and help the HEP community make the informed decision when choosing the portability solution to work with. 

The metrics are grouped according to the following categories and are documented here~\cite{metrics_doc}:

\begin{itemize}
    \item Ease of learning for experts and novices
    \item Ease of code conversion
    \begin{itemize}
        \item From CPU code to Accelerator (GPU, etc.) code
        \item From low level (CUDA, etc.) to higher level portability code
        \item From one portability framework to another
    \end{itemize}
    \item Impact on other existing code
    \begin{itemize}
        \item Extent of modifications to existing code: does it take over main(), does it affect the threading or execution model, etc.
        \item Extent of modifications to Event Data Model (EDM): data transfer and access across different memory space, etc.
    \end{itemize}
    \item Impact on existing tool chain and build infrastructure
    \begin{itemize}
        \item Extent of modifications to build rules / system
        \item Do we need to recompile the entire software stack?
        \item CMake or make changes/integration
    \end{itemize}
    \item Hardware mapping
    \begin{itemize}
        \item Is the technology working on all current hardware architectures?
        \item Support for new hardware features and new architectures
    \end{itemize}
    \item Feature availability
    \begin{itemize}
        \item Reductions, kernel chaining, callbacks, etc
        \item Concurrent kernel execution
        \item Support for interfacing to optimized math-heavy libraries (FFTs, etc.)
    \end{itemize}
    \item Ease of Debugging
    \begin{itemize}
        \item How easy is it to debug implementations of code in the technologies?
    \end{itemize}
    \item Address needs of all types of workflows
    \begin{itemize}
        \item Scaling with \# of kernels / application
        \item Scaling with \# of developers
        \item Support for users by portability technology developers
    \end{itemize}
    \item Long-term sustainability and code stability
    \begin{itemize}
        \item Support model of technologies, stability of implementation if underlying libraries (CUDA) change
        \item CUDA is going to be around for a long time, what about the portability solutions?
        \item Long term support for technologies by vendors
    \end{itemize}
    \item Compilation time
    \begin{itemize}
        \item Separate builds for different architectures?
        \item Compatibility with experiment’s software distribution strategies
    \end{itemize}
    \item Performance: CPU and GPU
    \begin{itemize}
        \item Does the portable code version (CPU and GPU uses same code) degrade the CPU performance or use more memory?
    \end{itemize}
    \item Aesthetics
    \begin{itemize}
        \item compatibility with C++ standards
    \end{itemize}
    \item Interoperability
    \begin{itemize}
        \item Can you mix portability technologies in the same application? How are external packages treated if they are imported into experiment software stacks and use different portability technologies? (CMSSW~\cite{jones:2006,jones:2014,jones:2015,jones:2017,bocci:2020a} is using Kokkos, but Geant~\cite{1610988,AGOSTINELLI2003250,ALLISON2016186} is using Alpaka)
        \item Interaction with existing thread pool on CPU/GPU back ends?
    \end{itemize}
\end{itemize}

\section{Use Cases}
\label{sec:usecases}
In the following, we briefly describe the use cases that are being used in this study.

\begin{description}
    \item[FastCaloSim] is a parametrized simulation of the ATLAS Liquid Argon Calorimeter~\cite{ATL-SOFT-PUB-2018-002}. The codebase was originally written in C++, then ported to CUDA. The CUDA implementation consists of 3 relatively small kernels, which perform a memory re-initialization, the main energy deposition simulation, and finally a stream compaction. It has been ported to Kokkos, SYCL, and std::par, targeting NVIDIA, Intel and AMD hardware.
    \item[ACTS] is a track reconstruction toolkit for general HEP detectors~\cite{andreas_salzburger_2022_6220148}, which is based on C++. The R\&D lines for ACTS parallelization on heterogeneous architectures consist of a number of core algorithms for tracking on GPUs (traccc), a geometry offloading package designed explicitly for GPUs (detray), and a memory management layer (vecmem) that is architecture neutral. All tracking algorithms, which include hit clusterization, seeding and Kalman filtering (both simple and combinatorial) will be offloaded onto the GPU to minimize the data transfers between host and device.
    \item[Wire Cell] The Wire-Cell Toolkit~\cite{wirecell_toolkit,Qian:2018qbv} is a C++ software library for the simulation, signal processing, reconstruction and visualization of Liquid Argon Time Projection Chamber (LArTPC) detectors for neutrino experiments, such as the planned Deep Underground Neutrino Experiments (DUNE~\cite{DUNE:2016hlj}). The use case we study is the LArTPC signal simulation module in Wire-Cell, which simulates the LArTPC detector response. So far the signal simulation module has been re-implemented in Kokkos~\cite{Dong:2022wxg}, and investigation with OpenMP is in progress. 
    \item[Patatrack] The Patatrack use case consists of CMS Heterogeneous Pixel Reconstruction~\cite{bocci:2020b,patatrack_standalone} code, that processes the raw pixel detector data up to pixel tracks and vertices, extracted into a standalone program. It includes a multi-threaded mock framework providing similar behavior as CMS software framework CMSSW~\cite{jones:2006,jones:2014,jones:2015,jones:2017,bocci:2020a}, and input data from CMS Open Data~\cite{ttbardata}. The original code was developed to run on NVIDIA GPUs with CUDA, accompanied with a simple translation header to allow compilation to CPU. We have ported the code to Kokkos and HIP, targeting NVIDIA and AMD GPUs. 
    \item[P2R] is a light-weight mini-app which performs the track propagation in radial direction and Kalman update kernels in track reconstruction~\cite{p2r}. With a simplified geometry and standalone setup, P2R can be used to test the performance of core tracking computation in various portability technologies in a shorter timescale. The original version is adapted from the mkFit project~\cite{mkfit}, which implements a parallel Kalman Filter Algorithm~\cite{kalman}, and has been re-implemented in CUDA, HIP, Kokkos, Alpaka, OpenACC and std::par.
    \item[Random Number Generators] We have leveraged the SYCL programming model and its interoperability with third-party libraries to demonstrate cross-platform performance portability across heterogeneous resources. We have implemented NVIDIA and AMD random number generator extensions to the oneMKL open-source interfaces library~\cite{onemkl}. The utility of our extensions are exemplified in a real-world setting via a high-energy physics simulation application, showing the performance of implementations that capitalize on SYCL interoperability are at par with native implementations, attesting to the cross-platform performance portability of a SYCL-based approach to scientific codes.
\end{description}

\section{Preliminary Results}
\label{sec:prelresults}
We gathered some preliminary results from our studies of various portability solutions.

\subsection{Kokkos}
Kokkos is a programming model and a C++ library for portable performance applications~\cite{kokkos}. It provides high-level parallel algorithms, such as for, prefix scan, and reduction, that can be nested with some restrictions, as well as multidimensional array data types. The mapping of work of the algorithms and the default layout of the multidimensional array depend on the chosen back end. Currently (Kokkos 3.5) these back ends include CPU serial, CPU parallel with OpenMP or Posix Threads, and device parallel with CUDA, HIP, HPX~\cite{Kaiser2020}, or SYCL. The high abstraction level on both algorithms and data is expected to provide reasonably good out-of-the-box performance also on computing architectures beyond CPUs and GPUs.

We were able to express all the custom algorithms in the use cases (FastCaloSim, Wire Cell Toolkit, Patatrack, p2r) in the Kokkos’ programming model, but we experienced some challenges. Kokkos requires its run time library to be built for one set of host serial, host parallel, and device parallel back end at a time, and e.g. in case of CUDA the library can support exactly one major GPU architecture version. While this approach works fine for HPC codes that are typically compiled for a specific supercomputer, it poses challenges for HEP experiment frameworks for which a single build is expected to be used in about 200 data centers with different computer hardware. In Kokkos 3.5, the CPU serial back end is thread safe, but it cannot be efficiently used from multiple threads, limiting its current usefulness in multi-threaded applications that process multiple collision events concurrently. Kokkos developers are working to improve the performance for this use case. Much of the HEP data is structured, and multidimensional arrays are useful only in limited use cases, implying a need for a separate solution for data structures. Currently Kokkos does not provide a unified, portable interface to Fast Fourier Transform (FFT) algorithms (e.g. to optimize platform-specific implementations), but such interface is being worked on.

For the WireCell toolkit use case, we have seen moderate performance speedups from multi-core CPUs, AMD GPUs and NVIDIA GPUs, using Kokkos. We have also demonstrated that running multiple concurrent processes to share the GPUs can further improve the performance gains, setting a promising direction for efficient utilization of HPC systems in Wire-Cell. For other use cases we saw performance degradation for specific hardware platforms (AMD GPUs in case of FastCaloSim).

\subsection{SYCL}
SYCL is an open-standard C++-based programming model that facilitates parallel programming on heterogeneous platforms. It provides a single source programming model, enabling developers to write both host-side and kernel code in the same file. Employing C++-based template programming, developers can leverage higher-level programming features in writing accelerator-enabled applications with the ability to integrate the native acceleration API, when needed, by using different interoperability interfaces provided by SYCL. The latest specification, SYCL 2020, is based on ISO C++17 standard, and features standard programming with templates and lambda functions to develop optimized code which can be offloaded to special purpose compute accelerators such as GPUs, FPGA, or AI/ML accelerators. The SYCL specification is designed to be a higher level abstraction above low-level native acceleration APIs with interoperability between existing libraries and other parallel programming models and can be built on top of OpenMP, Vulkan~\cite{vulkan}, OpenCL~\cite{opencl}, Kokkos, Raja~\cite{raja}, or some other back end. The SYCL programming model offers performance portability across various vendor hardware and interoperability with both open-source and closed-source (proprietary) software. As SYCL evolves, HPC-critical features will continue to be incorporated into the specification.

The applicability of our SYCL-based Random Number Generators (RNGs)~\cite{2021arXiv210901329P,9652858} has been evaluated in a GPU port of FastCaloSim~\cite{10.3389/fdata.2021.665783}. The interfaces we developed enabled the seamless integration of SYCL RNGs into FastCaloSim with no code modification across the evaluated platforms. The SYCL 2020 interoperability functionality enabled custom kernels and vendor-dependent library integration to be abstracted out from the application, leading to improvement of maintainability of the application and reducing the source lines of code. Using our RNG interfaces, we achieve comparable performance with native solutions on different architectures. Whereas the original C++ version of FastCaloSim had two separate code bases, x86 and CUDA, our RNG work has enabled event processing on a variety of major vendor hardware from a single SYCL entry point. Hence, the SYCL RNG based integration facilitates the code maintainability by reducing the FastCaloSim code size without introducing any significant performance overhead.
\subsection{OpenMP} 
OpenMP is a directive-based programming model that has evolved from a shared-memory programming model for multicore CPU architectures to one with rich features to support GPU accelerator offloading. The current version 5.1 of the OpenMP specification includes support for loop-level target offloading, memory management, and asynchronous CPU/GPU execution, all of which may be crucial for experimental HEP workflows. Major HPC hardware vendors such as AMD, HPE, Intel and NVIDIA are all onboard with the OpenMP model, and have been actively developing the compiler infrastructure to support the new and improved OpenMP features. However, as a directive-based programming model, it may not offer the same level of flexibility as language-based programming models such as Kokkos or SYCL. Nevertheless, the premise of only needing to add a few pragmas to the code and letting the compilers handle the low-level optimizations holds great promise for the future, when even more diverse and complex HPC architectures are to be expected. With the anticipated industrial support, OpenMP may become the portable programming model of choice in the future, and R\&D into how HEP software can take full advantage of this potential should be supported timely so that we don’t fall behind the rest of the scientific computing community. 

Our initial investigations into the OpenMP target offloading features have shown that in addition to the simple loop parallelism, many performance-enhancing features are also well supported and relatively easy to use. These examples include asynchronous kernel executions, interoperability with optimized vendor libraries and the ability to specify memory spaces.  Compiler support for OpenMP offloading has also been improving rapidly, with several functional compilers available on the market, such as the open-source LLVM Clang compiler, GNU C Compiler, HPE’s CCE compiler, along with NVIDIA’s, AMD’s and Intel’s compilers to support their own GPU architectures. While the performances with these compilers can vary quite a bit depending on the specific use case, they have all been steadily improving. 
\subsection{std::par}
std::parallel::execution (std::par) has been part of the C++ standard since C++17, and offers a high level interface to execute the contents of loops concurrently. Until recently, the concurrent back ends have been limited to the host side, using libraries such as TBB~\cite{tbb} to execute on different CPU threads and cores. Recently, both Intel and NVIDIA have released compilers that can target GPU devices (oneapi::dpl and nvc++ respectively). Given their high level nature, most low level functions and optimizations of domain specific languages such as CUDA and SYCL are not available, resulting in a loss of overall performance. However, the entry bar to users is extremely low, and requires little knowledge of GPU programming. Both of these compilers are still rather immature, exhibiting a number of compiler bugs and lack of build system integration, so performance numbers should be taken lightly. The continued development of these compilers is however a good indication that vendors are seeking standards based solutions, with both low and high level APIs.

\section{Conclusions}
The portable parallelization strategies (PPS) project of the High Energy Physics Center for Computational Excellence (HEP/CCE) is investigating solutions to the changing hardware architecture landscape of today and in the future. With accelerators like GPUs becoming more mainstream, especially in HPC systems, the development of scientific code is at a crossroad leaving the convenient era of x86-only code. To continue writing scientific code efficiently with a large and not always professionally trained user community to run on all hardware architectures, we think we need community solutions for portability techniques that will allow the coding of an algorithm once, and the ability to execute it on a variety of hardware products from many vendors. To that effect, the PPS project is investigating the feasibility of currently available portability solutions and will compare them based on a defined set of metrics. The goal is to develop recommendations for potential users when to deploy the appropriate solution. Preliminary results show deploying portability solutions is currently far from the convenience of having a standard, but there are encouraging successes using these solutions. We think without them, the scientific success of our experiments and endeavors is in danger, as software development could be expert driven and costly to be able to run on available hardware infrastructure. We think the best solution for the community would be an extension to the C++ standard with a very low entry bar for users, supporting all hardware forms and vendors. We are very far from that ideal though. In the future, as a community, we need to request and work on portability solutions and strive to reach this ideal.

\bibliographystyle{JHEP}
\bibliography{myreferences}  

\providecommand{\href}[2]{#2}\begingroup\raggedright\begin{thebibliography}{10}

\bibitem{cce_webpage}
``{HEP Center for Computational Excellence (HEP-CCE)}.''
  \url{https://www.anl.gov/hep-cce}.

\bibitem{cuda}
``{NVidia CUDA Toolkit}.'' \url{https://developer.nvidia.com/cuda-toolkit}.

\bibitem{sycl}
``{SYCL: C++ Single-source Heterogeneous Programming for OpenCL}.''
  \url{https://www.khronos.org/sycl/}.

\bibitem{hip}
``{HIP: C++ Runtime API and Kernel Language to create portable applications}.''
  \url{https://github.com/ROCm-Developer-Tools/HIP}.

\bibitem{Duarte_2018}
J.~Duarte, S.~Han, P.~Harris, S.~Jindariani, E.~Kreinar, B.~Kreis et~al.,
  \emph{Fast inference of deep neural networks in {FPGAs} for particle
  physics}, \href{https://doi.org/10.1088/1748-0221/13/07/p07027}{\emph{Journal
  of Instrumentation} {\bfseries 13} (2018) P07027}.

\bibitem{kokkos}
C.R.~Trott, D.~Lebrun-Grandié, D.~Arndt, J.~Ciesko, V.~Dang, N.~Ellingwood
  et~al., \emph{Kokkos 3: Programming model extensions for the exascale era},
  \href{https://doi.org/10.1109/TPDS.2021.3097283}{\emph{IEEE Transactions on
  Parallel and Distributed Systems} {\bfseries 33} (2022) 805}.

\bibitem{CarterEdwards20143202}
H.C.~Edwards, C.R.~Trott and D.~Sunderland, \emph{Kokkos: Enabling manycore
  performance portability through polymorphic memory access patterns},
  \href{https://doi.org/https://doi.org/10.1016/j.jpdc.2014.07.003}{\emph{Journal
  of Parallel and Distributed Computing} {\bfseries 74} (2014) 3202 }.

\bibitem{MathesP3MA2017}
A.~{Matthes}, R.~{Widera}, E.~{Zenker}, B.~{Worpitz}, A.~{Huebl} and
  M.~{Bussmann}, \emph{Tuning and optimization for a variety of many-core
  architectures without changing a single line of implementation code using the
  alpaka library},  Jun, 2017,
  \href{https://arxiv.org/abs/1706.10086}{https://arxiv.org/abs/1706.10086}
  [\href{https://arxiv.org/abs/1706.10086}{{\ttfamily 1706.10086}}].

\bibitem{ZenkerAsHES2016}
E.~Zenker, B.~Worpitz, R.~Widera, A.~Huebl, G.~Juckeland, A.~Kn{\"{u}}pfer
  et~al., \emph{Alpaka - an abstraction library for parallel kernel
  acceleration},  IEEE Computer Society, May, 2016,
  \href{http://arxiv.org/abs/1602.08477}{http://arxiv.org/abs/1602.08477}
  [\href{https://arxiv.org/abs/1602.08477}{{\ttfamily 1602.08477}}].

\bibitem{worpitz_2015_49768}
B.~Worpitz, \emph{{Investigating performance portability of a highly scalable
  particle-in-cell simulation code on various multi-core architectures}}, Ph.D.
  thesis, Technische Universität Dresden, Sept., 2015.
\newblock 10.5281/zenodo.49768.

\bibitem{dagum1998openmp}
L.~Dagum and R.~Menon, \emph{Openmp: an industry standard api for shared-memory
  programming}, {\emph{Computational Science \& Engineering, IEEE} {\bfseries
  5} (1998) 46}.

\bibitem{std_par}
``{ISO/IEC} 14882:2020: Programming languages -- {C++}.''
  \url{https://www.iso.org/standard/79358.html}, 2020.

\bibitem{:2008zzk}
{\scshape CMS} collaboration, \emph{{The CMS experiment at the CERN LHC}},
  \href{https://doi.org/10.1088/1748-0221/3/08/S08004}{\emph{JINST} {\bfseries
  3} (2008) S08004}.

\bibitem{DUNE:2016hlj}
{\scshape DUNE} collaboration, \emph{{Long-Baseline Neutrino Facility (LBNF)
  and Deep Underground Neutrino Experiment (DUNE)}: {Conceptual Design Report,
  Volume 1: The LBNF and DUNE Projects}},
  \href{https://arxiv.org/abs/1601.05471}{{\ttfamily 1601.05471}}.

\bibitem{Aad:2008zzm}
{\scshape ATLAS} collaboration, \emph{{The ATLAS Experiment at the CERN Large
  Hadron Collider}},
  \href{https://doi.org/10.1088/1748-0221/3/08/S08003}{\emph{JINST} {\bfseries
  3} (2008) S08003}.

\bibitem{andreas_salzburger_2022_6220148}
A.~Salzburger, P.~Gessinger, F.~Klimpel, M.~Kiehn, B.~Schlag, H.~G. et~al.,
  \emph{acts-project/acts: v17.1.0}, .

\bibitem{metrics_doc}
``{HEP-CCE Metrics for portability technologies}.''
  \url{https://hep-cce.github.io/Metric.html}.

\bibitem{jones:2006}
C.D.~Jones, M.~Paterno, J.~Kowalkowski, L.~Sexton-Kennedy and W.~Tanenbaum,
  \emph{The new {CMS} event data model and framework},  in \emph{Proceedings of
  International Conference on Computing in High Energy and Nuclear Physics
  (CHEP06)}, 2006.

\bibitem{jones:2014}
C.D.~Jones and E.~Sexton-Kennedy, \emph{Stitched together: Transitioning {CMS}
  to a hierarchical threaded framework},
  \href{https://doi.org/10.1088/1742-6596/513/2/022034}{\emph{J. Phys.: Conf.
  Series} {\bfseries 513} (2014) 022034}.

\bibitem{jones:2015}
C.D.~Jones, L.~Contreras, P.~Gartung, D.~Hufnagel and L.~Sexton-Kennedy,
  \emph{Using the {CMS} threaded framework in a production environment},
  \href{https://doi.org/10.1088/1742-6596/664/7/072026}{\emph{J. Phys.: Conf.
  Series} {\bfseries 664} (2015) 072026}.

\bibitem{jones:2017}
C.D.~Jones, \emph{{CMS} event processing multi-core efficiency status},
  \href{https://doi.org/10.1088/1742-6596/898/4/042008}{\emph{J. Phys.: Conf.
  Series} {\bfseries 898} (2017) 042008}.

\bibitem{bocci:2020a}
A.~Bocci, D.~Dagenhart, V.~Innocente, C.~Jones, M.~Kortelainen, F.~Pantaleo
  et~al., \emph{Bringing heterogeneity to the cms software framework},
  \href{https://doi.org/10.1051/epjconf/202024505009}{\emph{EPJ Web Conf.}
  {\bfseries 245} (2020) 05009}.

\bibitem{1610988}
J.~Allison, K.~Amako, J.~Apostolakis, H.~Araujo, P.~Arce~Dubois, M.~Asai
  et~al., \emph{Geant4 developments and applications},
  \href{https://doi.org/10.1109/TNS.2006.869826}{\emph{IEEE Transactions on
  Nuclear Science} {\bfseries 53} (2006) 270}.

\bibitem{AGOSTINELLI2003250}
S.~Agostinelli, J.~Allison, K.~Amako, J.~Apostolakis, H.~Araujo, P.~Arce
  et~al., \emph{Geant4—a simulation toolkit},
  \href{https://doi.org/https://doi.org/10.1016/S0168-9002(03)01368-8}{\emph{Nuclear
  Instruments and Methods in Physics Research Section A: Accelerators,
  Spectrometers, Detectors and Associated Equipment} {\bfseries 506} (2003)
  250}.

\bibitem{ALLISON2016186}
J.~Allison, K.~Amako, J.~Apostolakis, P.~Arce, M.~Asai, T.~Aso et~al.,
  \emph{Recent developments in geant4},
  \href{https://doi.org/https://doi.org/10.1016/j.nima.2016.06.125}{\emph{Nuclear
  Instruments and Methods in Physics Research Section A: Accelerators,
  Spectrometers, Detectors and Associated Equipment} {\bfseries 835} (2016)
  186}.

\bibitem{ATL-SOFT-PUB-2018-002}
{\scshape ATLAS Collaboration} collaboration, \emph{{The new Fast Calorimeter
  Simulation in ATLAS}},  Tech. Rep.
  \href{https://cds.cern.ch/record/2630434}{ATL-SOFT-PUB-2018-002}, CERN,
  Geneva (Jul, 2018).

\bibitem{wirecell_toolkit}
``Wire-cell toolkit.'' \url{https://github.com/WireCell/wire-cell-toolkit}.

\bibitem{Qian:2018qbv}
X.~Qian, C.~Zhang, B.~Viren and M.~Diwan, \emph{{Three-dimensional Imaging for
  Large LArTPCs}},
  \href{https://doi.org/10.1088/1748-0221/13/05/P05032}{\emph{JINST} {\bfseries
  13} (2018) P05032} [\href{https://arxiv.org/abs/1803.04850}{{\ttfamily
  1803.04850}}].

\bibitem{Dong:2022wxg}
Z.~Dong, K.~Knoepfel, M.~Lin, B.~Viren and H.~Yu, \emph{{Evaluation of Portable
  Programming Models to Accelerate LArTPC Detector Simulations}},  in
  \emph{{20th International Workshop on Advanced Computing and Analysis
  Techniques in Physics Research}: {AI Decoded - Towards Sustainable, Diverse,
  Performant and Effective Scientific Computing}}, 3, 2022
  [\href{https://arxiv.org/abs/2203.02479}{{\ttfamily 2203.02479}}].

\bibitem{bocci:2020b}
A.~Bocci, V.~Innocente, M.~Kortelainen, F.~Pantaleo and M.~Rovere,
  \emph{Heterogeneous reconstruction of tracks and primary vertices with the
  cms pixel tracker},
  \href{https://doi.org/10.3389/fdata.2020.601728}{\emph{Front. Big. Data}
  {\bfseries 3} (2020) 601728}
  [\href{https://arxiv.org/abs/2008.13461}{{\ttfamily 2008.13461}}].

\bibitem{patatrack_standalone}
``Standalone patatrack pixel tracking.''
  \url{https://github.com/cms-patatrack/pixeltrack-standalone/}.

\bibitem{ttbardata}
{CMS Collaboration}, ``{TTToHadronic\_TuneCP5\_13TeV-powheg-pythia8 in
  FEVTDEBUGHLT format for 2018 collision data. CERN Open Data Portal.}.''
  \href{http://doi.org/10.7483/OPENDATA.CMS.GOB0.0LEW}{doi:10.7483/OPENDATA.CMS.GOB0.0LEW},
  2019.

\bibitem{p2r}
``Light-weight mini-app which performs the track propagation in radial
  direction and kalman update kernels in track reconstruction.''
  \url{https://github.com/kakwok/p2r-tests}.

\bibitem{mkfit}
S.~Lantz, K.~McDermott, M.~Reid, D.~Riley, P.~Wittich, S.~Berkman et~al.,
  \emph{Speeding up particle track reconstruction using a parallel kalman
  filter algorithm},
  \href{https://doi.org/10.1088/1748-0221/15/09/p09030}{\emph{Journal of
  Instrumentation} {\bfseries 15} (2020) P09030–P09030}.

\bibitem{kalman}
R.E.~Kalman, \emph{{A New Approach to Linear Filtering and Prediction
  Problems}}, \href{https://doi.org/10.1115/1.3662552}{\emph{Journal of Basic
  Engineering} {\bfseries 82} (1960) 35}.

\bibitem{onemkl}
``{oneAPI Math Kernel Library (oneMKL) Interfaces}.''
  \url{https://github.com/oneapi-src/oneMKL}.

\bibitem{Kaiser2020}
H.~Kaiser, P.~Diehl, A.S.~Lemoine, B.A.~Lelbach, P.~Amini, A.~Berge et~al.,
  \emph{Hpx - the c++ standard library for parallelism and concurrency},
  \href{https://doi.org/10.21105/joss.02352}{\emph{Journal of Open Source
  Software} {\bfseries 5} (2020) 2352}.

\bibitem{vulkan}
``{Vulkan is a cross-platform industry standard enabling developers to target a
  wide range of devices with the same graphics API}.''
  \url{https://www.vulkan.org}.

\bibitem{opencl}
J.E.~Stone, D.~Gohara and G.~Shi, \emph{Opencl: A parallel programming standard
  for heterogeneous computing systems},
  \href{https://doi.org/10.1109/MCSE.2010.69}{\emph{Computing in Science
  Engineering} {\bfseries 12} (2010) 66}.

\bibitem{raja}
``{RAJA Performance Portability Layer}.'' \url{https://github.com/LLNL/RAJA}.

\bibitem{2021arXiv210901329P}
V.R.~{Pascuzzi} and M.~{Goli}, \emph{{Achieving near native runtime performance
  and cross-platform performance portability for random number generation
  through SYCL interoperability}}, {\emph{arXiv e-prints} (2021)
  arXiv:2109.01329} [\href{https://arxiv.org/abs/2109.01329}{{\ttfamily
  2109.01329}}].

\bibitem{9652858}
M.~Krainiuk, M.~Goli and V.R.~Pascuzzi, \emph{oneapi open-source math library
  interface},  in \emph{2021 International Workshop on Performance, Portability
  and Productivity in HPC (P3HPC)}, pp.~22--32, 2021,
  \href{https://doi.org/10.1109/P3HPC54578.2021.00006}{DOI}.

\bibitem{10.3389/fdata.2021.665783}
Z.~Dong, H.~Gray, C.~Leggett, M.~Lin, V.R.~Pascuzzi and K.~Yu, \emph{Porting
  hep parameterized calorimeter simulation code to gpus},
  \href{https://doi.org/10.3389/fdata.2021.665783}{\emph{Frontiers in Big Data}
  {\bfseries 4} (2021) }.

\bibitem{tbb}
``{oneAPI Threading Building Blocks}.''
  \url{https://github.com/oneapi-src/oneTBB}.

\end{thebibliography}\endgroup

\end{document}